\documentclass{aa}  

\usepackage{graphicx}
\usepackage{txfonts}
\usepackage{natbib}
\begin{document}

  \subtitle{Results from the CALYPSO IRAM-PdBI survey}

\title{Probing episodic accretion with chemistry: CALYPSO observations of the very-low-luminosity Class 0 protostar IRAM 04191+1522}
\titlerunning{Probing episodic accretion with chemistry: CALYPSO observations of IRAM 04191+1522}
  
  \author{S.~Anderl\inst{\ref{ipag}} \and
  S.~Maret\inst{\ref{ipag}} \and
  S.~Cabrit\inst{\ref{lerma}} \and
  A.~J.~Maury\inst{\ref{cea}} \and
  A.~Belloche\inst{\ref{mpifr}} \and
  Ph.~André\inst{\ref{cea}} \and
  A.~Bacmann\inst{\ref{ipag}} \and
  C.~Codella\inst{\ref{arcetri}} \and
  L.~Podio\inst{\ref{arcetri}} \and
  F.~Gueth\inst{\ref{iram}} 
}
      \authorrunning{S.Anderl et al.}

\institute{Univ. Grenoble Alpes, CNRS, IPAG, 38000 Grenoble,
  France\label{ipag}
  \and
  PSL Research University, Sorbonne Université,  Observatoire de Paris, LERMA, CNRS, 75014 Paris, France\label{lerma}
  \and
  AIM, CEA, CNRS, Université Paris-Saclay, Université Paris Diderot,
  Sorbonne Paris Cité, 91191 Gif-sur-Yvette, France\label{cea}
  \and
  Max-Planck-Institut für Radioastronomie, Auf dem Hügel 69, 53121
  Bonn, Germany\label{mpifr}
  \and
  INAF - Osservatorio Astrofisico di Arcetri,
  Largo E. Fermi 5, 50125 Firenze, Italy\label{arcetri}
  \and
  Institut de Radioastronomie Millimétrique (IRAM), 38406
  Saint-Martin-d'Hères, France\label{iram}
}

\date{Received 15 October 2019 / Accepted 15 September 2020}

 
  \abstract
%
   {The process of mass accretion in the earliest phases of star formation is still not fully understood: Does the accretion rate smoothly decline with the age of the protostar or are there short, intermittent  accretion bursts? The latter option would also yield the possibility for very low-luminosity objects (VeLLOs) to be precursors of solar-type stars, even though they do not seem to have sufficiently high accretion rates to reach stellar masses during their protostellar lifetime. Nevertheless, probing such intermittent events in the deeply embedded phase is  not easy. Chemical signatures in the protostellar envelope can trace a past accretion burst.
}
   {We aim to explore whether or not the observed C$^{18}$O and N$_2$H$^+$ emission pattern towards the VeLLO IRAM 04191+1522 can be understood in the framework of a scenario where the emission is chemically tracing a past accretion burst.
}
   {We used high-angular-resolution Plateau de Bure Interferometer (PdBI) observations of C$^{18}$O and N$_2$H$^+$ towards IRAM 04191+1522 that were obtained as part of the CALYPSO IRAM Large Program (Continuum And Lines in Young ProtoStellar Objects). We model these observations using a chemical code with a time-dependent physical structure coupled with a radiative transfer module, where we allow for variations in the source luminosity.
   }
   {We find that the N$_2$H$^+$ line emission shows a central hole, with the N$_2$H$^+$ emission peaking at a radius of about 10$\arcsec{}$ (1400 au) from the source, while the C$^{18}$O emission is compact (1.3$\arcsec{}$ FWHM, corresponding to 182 au). The morphology of these two lines cannot be reproduced with a constant luminosity model based on the present-day internal luminosity (0.08 L$_{\sun}$). However, the N$_2$H$^+$ peaks are consistent with a constant-luminosity model of 12 L$_{\sun}$. Using a model with time-dependent temperature and density profiles, we show that the observed N$_2$H$^+$ peak emission could indeed be caused by a past accretion burst with a luminosity 150 times higher than the present-day luminosity. Such a burst should have occurred a couple of hundred years ago.}
   {We suggest that an accretion burst occurred in IRAM 04191+1522 in the recent past. If such bursts are common and sufficiently long in VeLLOs, they could lead to higher accretion onto the central object than their luminosity suggests. For IRAM 04191 in particular, our results yield an estimated final mass of 0.2 - 0.25 M$_{\sun}$ by the end of the Class 0 phase, which would make this object a low-mass star rather than a brown dwarf. More generally, our analysis demonstrates that the combination of observations of N$_2$H$^+$ and C$^{18}$O is a more reliable diagnostic of past outburst activity than C$^{18}$O or N$_2$H$^+$ emission alone.}


   \maketitle
%

\section{Introduction }

Very young, deeply embedded protostars with a stellar mass still smaller than the envelope mass are in the phase of heavy accretion, building up their final mass. The details of how these Class 0 protostars obtain their mass remain poorly understood, even though this phase is crucial for the subsequent evolution of sun-like stars and sets the initial conditions for the establishment of their final properties, such as their final mass or that of their protoplanetary disk \citep[e.g., ][]{Andre:1993}. In particular, it is unclear whether mass accretion can be understood as a constant or smoothly declining process, or whether or not there are intermittent periods of high-accretion bursts. A hint towards the latter option may be given by the so-called "luminosity problem" \citep[see, e.g.,][]{Kenyon:1990,Kenyon:1994,Kenyon:1995,Offner:2011,Dunham:2013}: Observational estimates of accretion rates based on protostellar bolometric luminosities are much lower than mass-accretion rates as derived from theoretical collapse models.

One special aspect of this problem concerns the existence of very-low-luminosity objects \citep[VeLLOs; e.g. ][]{DiFrancesco:2006}. These VeLLOs were discovered with {\it Spitzer} in its search for starless cores at the onset of accretion and are defined as embedded objects with an internal luminosity (i.e., luminosity from the star and a possible disk) of $L_\mathrm{int}$ $\le$ 0.1 L$_{\sun}$. Based on the standard collapse model \citep{Shu:1977}, which predicts spherical mass accretion at a rate of $\dot{M}_{\rm acc}\sim$2$\times$10$^{-6}$ M$_{\sun}$yr$^{-1}$, for a protostar with a typical radius of $R\sim$ 3~R$_{\sun}$ to reach the stellar/substellar boundary ($M$= 0.08 M$_{\sun}$) an accretion luminosity of at least 1.6~L$_{\sun}$ is expected \citep[e.g., ][]{Dunham:2006}. If a systematic underestimation of the observed luminosities of VeLLOs can be excluded, this means that either more mass is accreted during short outbursts where the accretion is temporarily much higher than currently observed, or that the final mass is much less than stellar, and the object becomes a brown dwarf. It is therefore of high theoretical importance to understand how to observationally trace such past accretion bursts if they indeed occur.

One idea is to use the chemistry in the protostellar envelope as such a tracer. This idea relies on the fact that characteristic chemical timescales are usually long enough for the envelope to still show the chemical imprint of a past burst, even when the burst is already over. For instance, in the cold and dense central part of protostellar envelopes, the CO ice depletion timescale is on the order of 10,000 yr. Such imprints have been observed towards more evolved sources with a known past burst, such as for example the T-tauri star EX Lup \citep{Abraham:2009,Banzatti:2012}. In deeply embedded sources, the burst-induced change in their envelope temperature profile can affect the deuteration fraction of certain species \citep{Owen:2015} and sublimate some ices from the grains in a larger region than expected from its post-burst, quiescent state luminosity \citep{Visser:2015}. Using the sublimation of ices, observational studies so far have mostly focused on C$^{18}$O as a tracer of the CO snow line and compared its radius to the one theoretically expected from the source's present luminosity \citep{Jorgensen:2015,Anderl:2016,Frimann:2017}. In this paper, we use N$_2$H$^+$ (together with
C$^{18}$O) as another tracer of a past accretion burst \citep[see ][]{Hsieh:2019,Hsieh:2018,vantHoff:2017} from observations towards the Class 0 protostar IRAM 04191+1522.

IRAM 04191+1522 (abbreviated here to IRAM~04191) is a low-luminosity protostar in the Taurus molecular cloud at a distance of $d$ = 140 pc \citep{Maheswar:2011,Zucher:2019}. \citet{Maury:2019} showed it to be a singular source and thus refuted a possible binary nature as claimed by \citet{Chen:2012}. The corresponding continuum maps of IRAM~04191 at 231 and 94 GHz from our CALYPSO dataset are shown in Figure B.6 of that paper. It was discovered through millimeter dust continuum mapping by \citet{Andre:1999} and was identified as a very young accreting Class 0 protostar based on its very high submillimeter-to-bolometric luminosity ratio of $\sim$12\% and its very low bolometric temperature of $\sim$18 K. Its bolometric luminosity was observed as 0.15 L$_{\sun}$. A rather high envelope mass of $\sim$0.5 M$_{\sun}$ is estimated within a radius of 4200 au, while the central mass is estimated as being a factor ten lower than that. 
\citet{Andre:1999} already report a well-defined, collimated CO bipolar outflow. The PA of its axis projected onto the plane of the sky was determined by \citet{Belloche:2002} as 28$^\circ$. IRAM~04191 was classified as a VeLLO by \citet{Dunham:2006} using {\it Spitzer} observations, which hinted at an internal luminosity of $L_{\rm int}$ = 0.08~L$_{\sun}$\footnote{A slightly lower value of the internal luminosity was derived from the analysis of Herschel maps obtained in the framework of the Gould Belt survey (0.05~L$_{\sun}$, Ladjelate et al. in prep.).}. 

Our observations were obtained as part of the IRAM PdBI CALYPSO survey (Continuum And Lines in Young Proto-Stellar Objects\footnote{See the project page at \url{http://irfu.cea.fr/Projets/Calypso} and the IRAM archive page at \url{https://www.iram-institute.org/EN/content-page-317-7-158-240-317-0.html}}). This survey studies different aspects of the earliest phases of low-mass star formation, such as outflow physics, envelope chemistry, protostellar disks, and multiplicity based on Plateau de Bure interferometer (PdBI) observations towards a large sample of the nearest Class 0 protostars \citep{Maret:2014,Maury:2014,Codella:2014,Santangelo:2015,Anderl:2016,Podio:2016,Maury:2019}.

The present paper presents a follow-up study of \citet{Anderl:2016}, where we probed the CO snow line in the whole CALYPSO sample of young protostars using the emission of C$^{18}$O and N$_2$H$^+$. Eleven out of sixteen sources showed anti-correlated emission of these two tracers. This anti-correlation can be traced back to N$_2$H$^+$ destruction by CO, once CO is sublimated from the dust at elevated temperatures close to the protostar \citep{Bergin:2001,Maret:2006}. \citet{Anderl:2016} modeled the chemistry underlying this emission for four sources (IRAS4B, IRAS4A, L1448C, and L1157) that showed a ring-like structure in N$_2$H$^+$ around the protostar and revealed the need for a higher sublimation temperature of CO ices than the value for pure CO ices (24 K instead of 19 K).

IRAM~04191 was found to have an emission pattern of C$^{18}$O and N$_2$H$^+$  that is  inconsistent with the explanation suggested for the other four sources, as its emission peaks in N$_2$H$^+$ do not adjoin to the emission in C$^{18}$O as in the other sources. The central hole of N$_2$H$^+$ emission towards IRAM~04191 was first observed and discussed by \citet{Belloche:2004} using lower angular resolution PdBI and IRAM 30m data. These latter authors tentatively attributed this central emission hole to N$_2$H$^+$ depletion towards the source center. In order to reconcile this interpretation with chemical modeling, which did not predict such a depletion, they suggested the assumption of a higher N$_2$ binding energy than the usually
adopted value of 750 K. Alternatively, they suggested the lack of N$_2$H$^+$ could be accounted for by enhanced deuteration of N$_2$H$^+$ in that region.

In this article, we aim to investigate whether or not the observed chemical envelope morphology in IRAM~04191 can be understood using chemical modeling that allows for the possibility of a past accretion burst. The paper is structured as follows: In Section \ref{Observations}, we present our observations, which we evaluate in Section \ref{Morphology}. Section \ref{Analysis} then presents a detailed analysis based on our modeling approach, where three different scenarios with respect to the source luminosity are
probed. The results are discussed in Section \ref{Discussion} and our conclusions are given in Section \ref{Conclusions}.

\section{Observations}\label{Observations}

Observations towards IRAM~04191 ($\alpha_{\rm J2000}$ $=$ 04$^\mathrm{h}$21$^\mathrm{m}$56$\fs$903, $\delta_{\rm J2000}$ $=$ 15$\degr$29$\arcmin$46$\farcs$15) were performed with the PdBI between November 2010 and November 2011 using the A and C configurations of the array (baselines between 19 and 762 m). The lines were observed in two different spectral setups, each using the narrow-band backend. For the C$^{18}$O (2--1) line at 219.560\,354 GHz (1.4 mm), this backend provided a bandwidth of 250 channels of 73 kHz (0.10 km s$^{-1}$) each, while for the N$_2$H$^+$ (1--0) hyperfine lines, whose brightest component is centered at 93.173\,770 GHz (3.2 mm), the bandwidth corresponds to 512 channels of 39 kHz (0.13 km s$^{-1}$). Calibration was done using \texttt{CLIC}, which is part of the \texttt{GILDAS} software suite\footnote{\url{http://www.iram.fr/IRAMFR/GILDAS/}}. For the 1.4 mm observations, the phase root mean square (RMS) was $<$ 65$^{\circ}$, with precipitable water vapor (PWV) between 0.7 mm and 2.0 mm, and system temperatures ($T_{\rm sys}$) $<$ 150 K. All data with phase rms $>$ 50$^{\circ}$ were flagged before producing the uv tables. For the 3.2 mm observations, the phase RMS was $<$ 50$^{\circ}$, with PWV between 0.9 mm and 3 mm and $T_{\rm sys}$ $<$ 80 K. Here, all data with phase RMS $>$ 40$^{\circ}$ were flagged before producing the uv tables. The continuum emission was removed from the visibility tables to produce continuum-free line tables. Spectral datacubes were produced from the visibility tables using a natural weighting, and deconvolved using the standard \texttt{CLEAN} algorithm in the \texttt{MAPPING} software. The RMS noise in the final velocity-integrated datacubes is 13.0 mJy beam$^{-1}$ km s$^{-1}$ for C$^{18}$O and 14.7 mJy beam$^{-1}$ km s$^{-1}$ for N$_2$H$^+$. The synthesized beam sizes are 0.9$''$ $\times$ 0.7$''$ (130 au $\times$ 100 au) for C$^{18}$O and 1.9$''$ $\times$ 1.6$''$ (270 au $\times$ 220 au) for N$_2$H$^+$.

\section{Results}\label{Morphology}
\begin{figure}
  \includegraphics[width=\columnwidth]{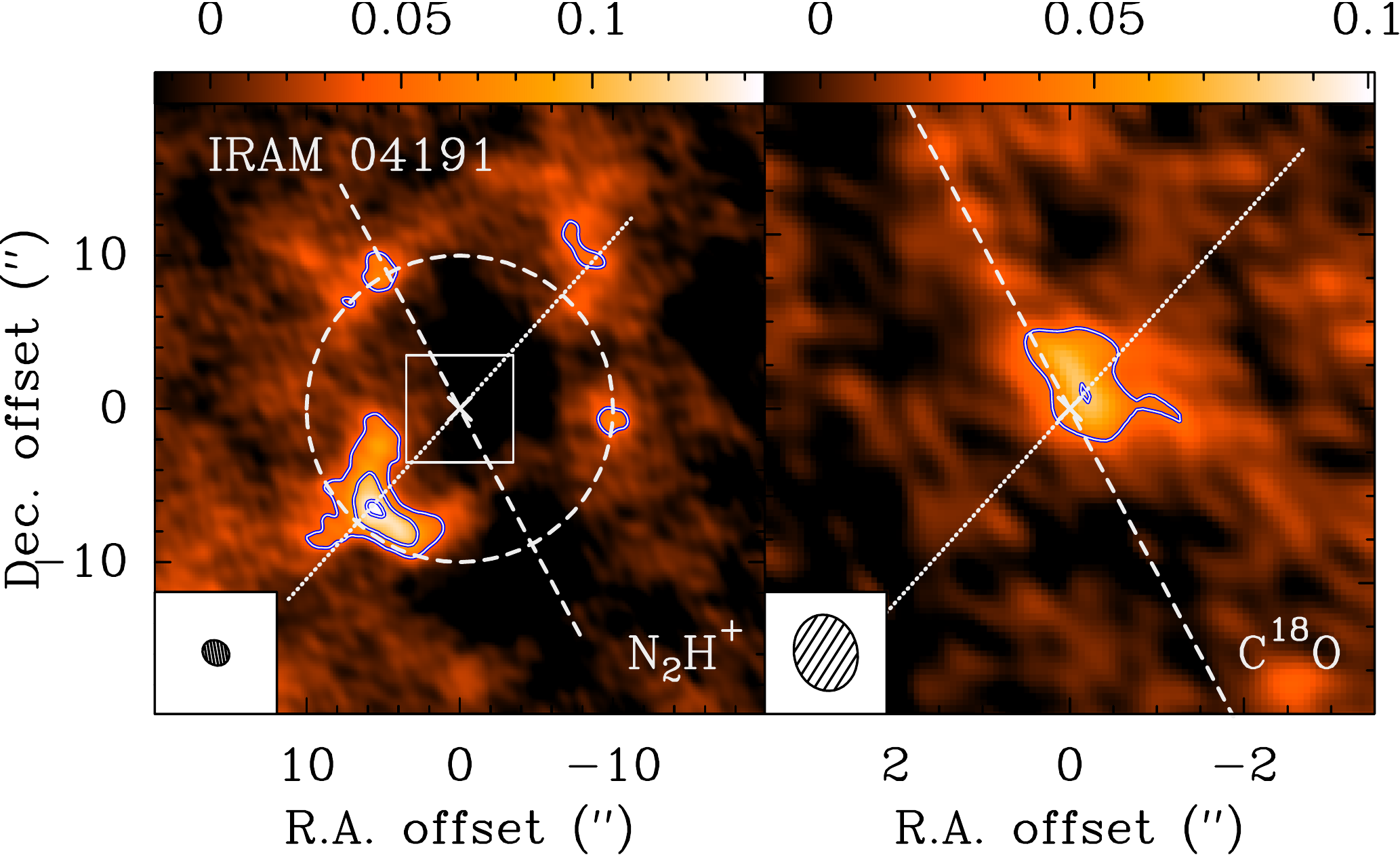}
  \caption{Left panel: N$_2$H$^+$ (1--0) emission integrated over all seven hyperfine components.  Contours show the emission in steps of 3$\sigma$, starting at 3$\sigma$, with $\sigma$=15 mJy beam$^{-1}$ km s$^{-1}$.
    The filled ellipse
    in the lower left corner of the large panel indicates the synthesized
    beam size of the N$_2$H$^+$ observations at 3 mm, while the FWHM primary beam is 
    52$''$ at 93 GHz. The dashed white
    line illustrates the outflow direction
    \citep[PA = 28$^\circ$,][]{Belloche:2002}, and the white cross shows the position
    of the continuum source as determined in \citet{Maury:2019} (see their Fig. B.6). The dotted line shows the
    direction of the cuts through the N$_2$H$^+$ emission maximum
    shown in Fig. \ref{observations_cut}. The dashed circle marks a radius of
    10$''$, which was stated as being the N$_2$H$^+$ hole rim radius by
    \citet{Belloche:2004} and \citet{Lee:2005}.  Right panel: Zoom into the inner $\pm$2$''$ in integrated
    C$^{18}$O emission. Contours are in steps of 3$\sigma$, starting at 3$\sigma$, with $\sigma$=13 mJy beam$^{-1}$ km s$^{-1}$. The corresponding synthesized beam size at 1.4 mm is displayed in the lower left corner, while the FWHM primary beam is 22$''$ at 220 GHz. The white lines are the same as in the left panel. \label{observations_map}}
\end{figure}

Figure \ref{observations_map} shows the cleaned maps of C$^{18}$O (2-1) and N$_2$H$^+$ (1-0) as observed with the PdBI. The observations in C$^{18}$O are integrated over $\pm$3 km s$^{-1}$ around the systemic velocity, which is 6.7 km s$^{-1}$ for IRAM~04191 \citep{Belloche:2002,Takakuwa:2003}. The N$_2$H$^+$ emission is integrated over a continuous window of 20 km s$^{-1}$ which covers all hyperfine structure components of the (1-0) transition.

We detect very compact emission in C$^{18}$O (2-1) towards the source continuum peak, with a maximum velocity integrated intensity of about 6$\sigma$. An elliptical Gaussian fit of the integrated data in the uv plane with fixed central coordinates towards the continuum source position yields an extent of (1.4 $\pm$ 0.3)$''$ $\times$ (1.3 $\pm$ 0.3)$''$ at a PA of -62 $\pm$ 93 $^\circ$. This marked emission peak is still clearly visible if we combine the interferometer data with IRAM 30m single dish data (see Appendix~\ref{shortspacing}, Figures \ref{map_shortspacing} and \ref{cuts_shortspacing}), which add the information on scales larger than about 9$''$ for CO and 21$''$ for N$_2$H$^+$. Accordingly, in the combined C$^{18}$O data, we see two components: one broad pedestal and the narrow peak on top. The broad component seems to stem from the extended envelope around the IRAM~04191 core. As we are interested in the emission originating from the abundance enhancement in the innermost envelope close to the central source, we focus on the narrow component traced in our PdBI data in the following analysis. Furthermore, it is noteworthy that the drop in C$^{18}$O emission at a FWHM of about 1.3$''$ is a feature occurring on scales fully probed by the PdBI data.

The integrated emission in N$_2$H$^+$ (1-0) shows a central hole, which has already been observed and discussed by \citet{Belloche_2004} and \citet{Lee:2005}. In the shortspacing-corrected data, we again see two different components: an extended component stemming from the flattened envelope at scales larger than 21$''$ and the central
depression framed by emission peaks (see Figure \ref{cuts_shortspacing}). Most importantly, the shortspacing-corrected data show the same location for these N$_2$H$^+$ emission peaks as the uncorrected data. As the peak location is the piece of information that we base our analysis on, for N$_2$H$^{+}$ we  also focus on the emission in the innermost envelope, which is traced by the PdBI data. The distance of the peaks from the center is about 10$''$ as most easily measured on the bright emission arc southeast of the source. The N$_2$H$^+$ peaks do not attach to the emission in C$^{18}$O as would be expected if the maxima were due to actual N$_2$H$^+$ destruction by CO and as was observed in four other CALYPSO sources \citep{Anderl:2016}. Spectra of CO and N$_2$H$^+$ at the three most significant positions are shown in the Appendix.~\ref{sec:n_2h+-c18o-spectra}

\begin{figure}
  \includegraphics[width=\columnwidth]{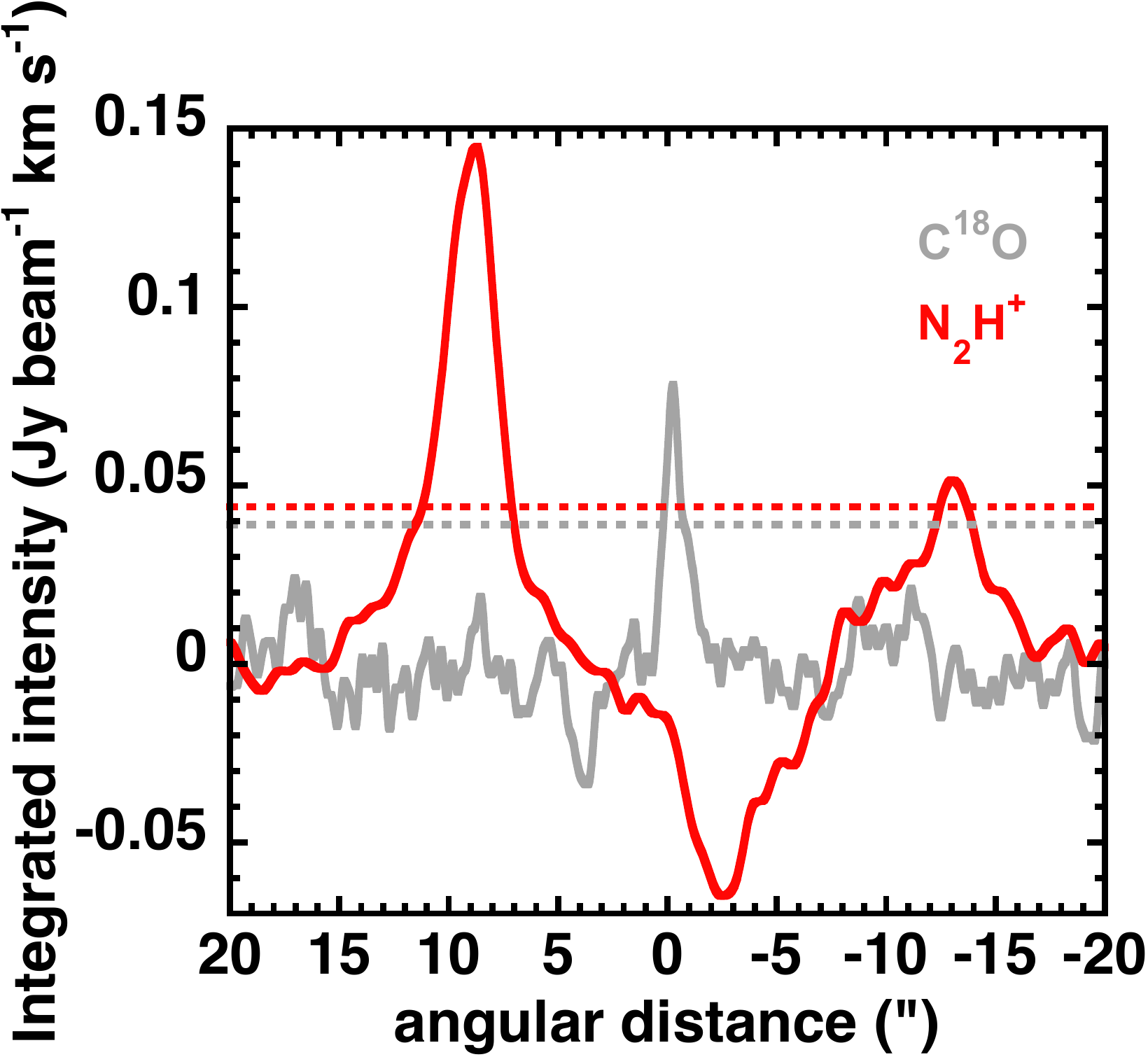}
  \caption{Cuts in the image plane through the southeastern
    N$_2$H$^+$ emission peak (PA = 138$^\circ$) for C$^{18}$O (gray)
    and N$_2$H$^+$ (red) with corresponding 3$\sigma$-levels (dotted lines).} \label{observations_cut}
\end{figure}

Figure \ref{observations_cut} shows cuts in the image plane through the southeastern N$_2$H$^+$ emission peak at PA = 138$^\circ$. The very narrow, compact C$^{18}$O (2-1) peak is clearly visible, as are the two peaks in N$_2$H$^+$ (1-0) emission. The negative N$_2$H$^+$ emission in the cut is due to the spatial filtering of the extended pedestal seen in images with the short-spacing correction (see Appendix~\ref{shortspacing}).

\section{Analysis}
\label{Analysis}
\subsection{The model}

\begin{table*}
  \caption{Model parameters and properties.\label{modelparam}}
  \centering                          
  \begin{tabular}{lllllllll}
    \hline
    \hline
    Model  & & \multicolumn{2}{c}{Parameters} & & \multicolumn{4}{c}{Properties}\\
    \cline{3-4}
    \cline{6-9}
                   &  & X(C$^{18}$O)$_{\rm init}$\tablefoottext{a} & X(N$_2$)$_{\rm init}$\tablefoottext{b} &  & X(C$^{18}$O)$_{\rm plateau}$\tablefoottext{c} & X(N$_2$H$^+$)$_{\rm peak}$\tablefoottext{d} & r$_{\rm snow}$(CO)\tablefoottext{e} & r$_{\rm snow}$(N$_2$)\tablefoottext{f} \\
                   &  &                           &                       &  &                              &                            & (au)               & (au)                  \\
    \hline   
    Model1         &  & 1.8$\times$10$^{-8}$      & 2$\times$10$^{-7}$    &  & 1.4$\times$10$^{-8}$         & 2$\times$10$^{-10}$        & 70                 & 130                   \\
    Model2         &  & 1.3$\times$10$^{-8}$      & 2$\times$10$^{-7}$    &  & 1.3$\times$10$^{-8}$         & 2$\times$10$^{-10}$        & 1100               & 2000                  \\
    Time dependent &  & 1.4$\times$10$^{-8}$      & 2$\times$10$^{-7}$    &  & 1.4$\times$10$^{-8}$         & 2$\times$10$^{-10}$        & -                  & -                     \\

    \hline           
  \end{tabular}
  \tablefoot{
  \tablefoottext{a}{Initial fractional abundance of C$^{18}$O. All fractional abundances in this table are given relative to H$_2$.}
  \tablefoottext{b}{Initial fractional abundance of \textbf{N$_2$}.}
  \tablefoottext{c}{Central plateau fractional abundance of C$^{18}$O.}
  \tablefoottext{d}{Peak fractional abundance of N$_2$H$^+$.}
  \tablefoottext{e}{Radius of the CO snow line.}
  \tablefoottext{f}{Radius of the \textbf{N$_2$} snow line.}}
\end{table*}

In order to model the observed emission in C$^{18}$O and N$_2$H$^+$,
we use the same modeling approach as in \citet{Anderl:2016} (a detailed description of the model and its parameters is found there): Based on the dust temperature and density profiles of the source, we first calculate one-dimensional chemical abundance profiles with the \texttt{astrochem} 1D chemistry code \citep{Maret:2015} as a function of time. For that, we assume the gas and dust to have the same temperature, which is justified at the high densities considered (between $\sim$5$\times$10$^5$~cm$^{-3}$ and 2$\times$10$^8$~cm$^{-3}$). The code can be applied to stationary source density and temperature profiles as well as to profiles that change in time. Each shell of the source is at constant temperature and density. We use the same input parameters and general initial conditions for the chemistry code as in \citet{Anderl:2016}. In particular, we use a CO binding energy of 1200 K (corresponding to a sublimation temperature of $T_{\rm sub}$ = $\sim$24 K) and a N$_2$ binding energy of 1000 K ($T_{\rm sub}$ = $\sim$19 K), which were shown to reproduce the observed emission morphologies in all four sources modeled in this latter study (see also \citealt{Yildiz:2012,Fayolle:2016}). Having thus constrained these values in the previous study, we leave them fixed here and only vary the initial CO and N$_2$ ice abundances in order to arrive at similar peak intensities as observed. The initial abundances for C$^{18}$O and N$_2$ are listed in Table \ref{modelparam}.

We feed the resulting abundance profiles into the radiative transfer code \texttt{RATRAN} \citep{Hogerheijde:2000}. Based on the Monte Carlo method, this code computes the molecular excitation and the resulting molecular line emission for an axially symmetric source model. We used the collisional rates for C$^{18}$O and N$_2$H$^+$ as provided by the Leiden Atomic and Molecular Database \citep{Yang:2010,Daniel:2005}. As in our previous study, from the resulting images we finally compute uv tables with the \texttt{GILDAS} mapping software using the same uv coverage as in our observations. The observed and modeled uv data are then processed with the same imaging and cleaning steps.

While in \citet{Anderl:2016} we used the source temperature and density profiles provided by \citet{Kristensen:2012}, here we calculate the source temperature profile based on the source internal luminosity and an assumed 1D power-law density profile using the radiation transfer code \texttt{Transphere}. This code solves the 1D dust radiative transfer by applying the variable Eddington factor method based on absorption and re-emission processes, using the dust opacity of 0.1 $\mu$m-sized silicate grains \citep{Draine:1984}. The details of the code are described in \citet{Dullemond_2002}. This change in the modeling strategy allows us to simulate luminosity burst scenarios using the burst luminosity as a free input parameter, similar to the approach applied in \citet{Jorgensen:2015}. We use the density profile presented in \citet{Belloche:2002} as input.

\subsection{Constant low-luminosity model}

\begin{figure*}
  \centering
  \includegraphics[width=17cm]{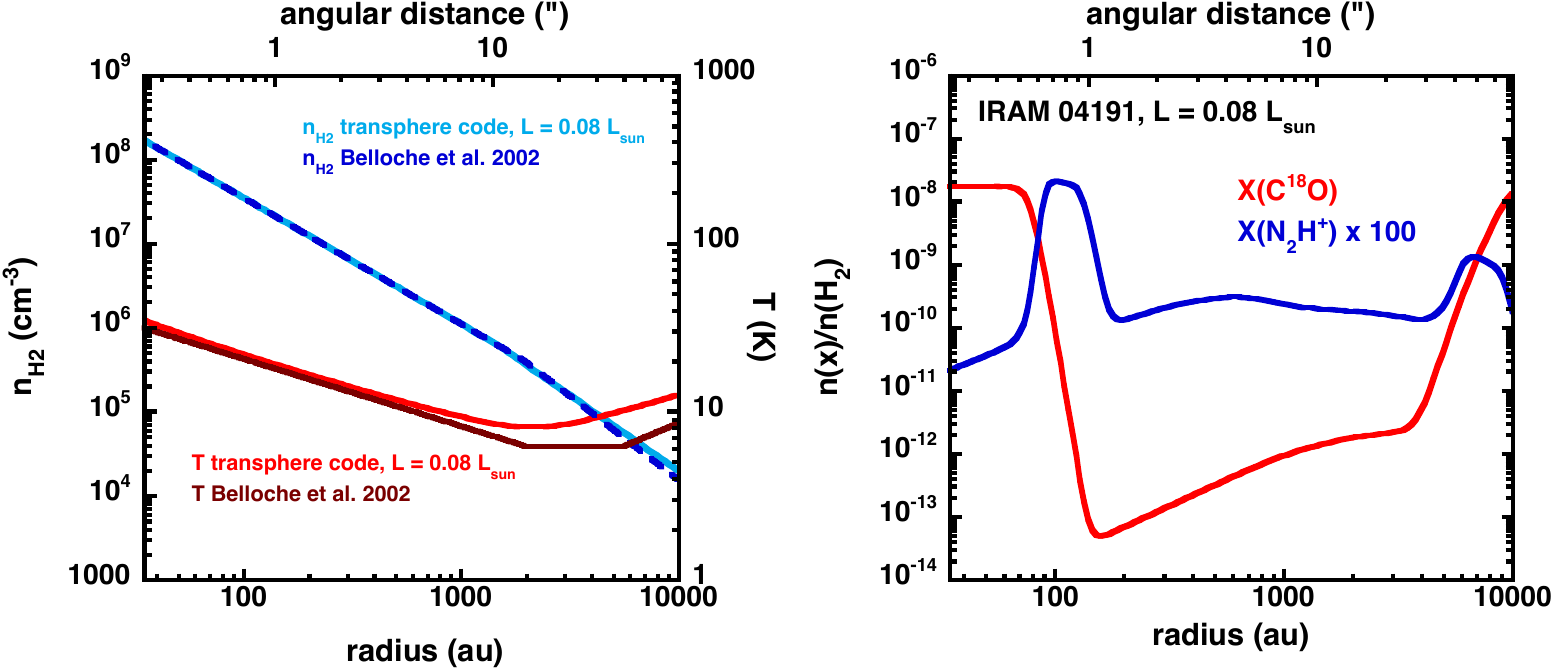}
  \caption{Constant low-luminosity model (Model1). {\it Left:} Density
    and dust temperature profiles of IRAM~04191 as presented in Belloche et
    al. 2002 (temperature: brown, density: dashed dark blue) and as
    self-consistently computed with the transphere code for an
    internal luminosity of 0.08 L$_{\sun}$ (temperature: red, density:
    light blue). {\it Right:} Chemical abundances of C$^{18}$O (red)
    and N$_2$H$^+$ (blue, upscaled by a factor 100) at a chemical age
    of $\sim$ 10$^4$ yr. \label{model1_prof_chem}}
\end{figure*}

In a first step, we try to reproduce the observed emission morphologies in C$^{18}$O and N$_2$H$^+$ based on the current internal luminosity value of 0.08 L$_{\sun}$ (we refer to this model as model1). Following \citet{Belloche:2002}, we use a density profile consisting of a piecewise power-law function $\rho \propto r^{-p}$ with an exponent $p = 1.5$ inside the innermost 1500 au and $p = 1.8$ outside of that radius. The power-law density index $p$ outside of a radius of 1500 au stems from the "best-fit" index of the averaged radial intensity profile from IRAM 30m 1.3 mm continuum observations with the assumption of an outwards increasing temperature profile for this low-luminosity protostar due to external heating \citep{Motte:2001}. Accordingly, as the interstellar radiation field is assumed to have an effect on the temperature structure due to the source's low internal luminosity of 0.08~L$_{\sun}$, we account for the external radiation by adding five components (three in V-NIR, one FIR, and one CBR) as described in \citet{Zucconi_2001}. Based on these inputs, we then calculate self-consistently the temperature profile using the \texttt{transphere} continuum radiative transfer code. The resulting temperature profile lies slightly above the piece-wise power-law temperature profile used by Belloche et al. 2002 (by $\sim$5\% between radii of 30 and 1000 au, see Fig. \ref{model1_prof_chem}), which was constrained by molecular line observations. We note that these authors did not self-consistently calculate the temperature profile, and therefore a slight deviation is expected. Our profile however matches their observational constraints derived from C$^{18}$O and N$_2$H$^+$ observations, which is a temperature of 10 $\pm$ 2 K in the low-density outermost part of the envelope ($r$ $\sim$ 6000 au) and temperatures higher than $\sim$6 K in the N$_2$H$^+$-emitting part of the envelope. Our temperature in the regime of radii between 2000 au and 6000 au is however significantly higher than their estimate of $\sim$6-7 K as derived from CS observations (between 8.5 K and 10 K in our profile), while the temperature in the outermost part of the envelope is 12 K instead of 10 K, which is the typical temperature of the Taurus cloud \citep{Benson:1989}. The reason for this is that we did not fine-tune the assumed interstellar radiation field to the particular environment of IRAM~04191 because we are mostly interested in the innermost part of the envelope within radii of $\sim$2000 au. Accordingly, the small differences between both profiles in the outermost envelope does not have an impact on our  analysis.

\begin{figure*}
  \sidecaption
  \includegraphics[width=12cm]{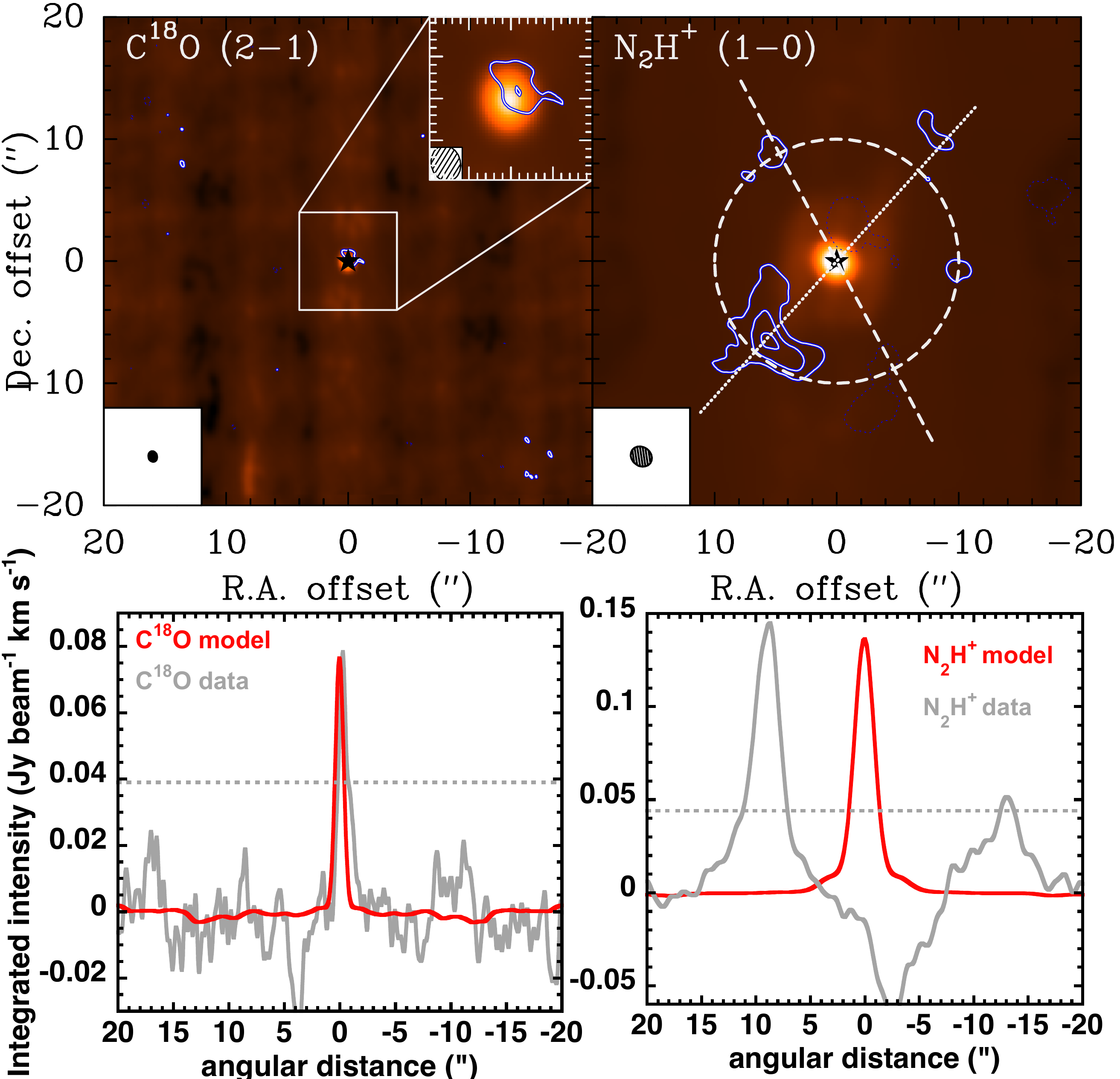}
  \caption{Constant low-luminosity model (Model1). {\it Top row:}
    Comparison of the observations (contours) with the synthetic maps
    (color background). Top left: C$^{18}$O (2--1). The inlay in the upper right corner of the left map
    shows a zoom into the C$^{18}$O emission within the central
    2$''$. Right: N$_2$H$^+$
    (1--0); integration intervals are the same as in Fig. 1, while the
    contour spacing is in steps of 3$\sigma$, starting at
    3$\sigma$. The dotted line shows the direction of the cuts through the maps, while the dashed line illustrates the outflow direction. The white dashed circle marks a radius of 10$''$. The scaling of the maps can be understood in comparison
    with the cuts in the bottom row. {\it Bottom row:} Cuts at PA =
    138$^\circ$ for C$^{18}$O (left) and N$_2$H$^+$ (right). Gray
    lines show the data, red lines the model, and the dotted gray lines show the 3$\sigma$. \label{model1}}
\end{figure*}

Based on the density and temperature profiles, we then computed the
envelope chemistry using the \texttt{astrochem} code (a more detailed
description of this step is found in \citealt{Anderl:2016}). We choose
a chemical age (i.e., time after which we stop the computation) of $\sim$10$^4$ yr in accordance with the maximum age
estimated by \citet{Belloche:2002}, which is 3-5 $\times$
10$^4$ yr from accretion and envelope-dissipation scenarios. Figure
\ref{model1_prof_chem} shows the resulting chemical abundances of
C$^{18}$O and N$_2$H$^+$, which show the expected anticorrelation
between both species within the central $\sim$100 au: C$^{18}$O is present in the gas phase only in
the innermost part of the envelope, where the temperature is higher
than the CO freeze-out temperature, determined to be $\sim$ 24 K by
\citet{Anderl:2016}. Once it is in the gas phase, CO
proton-transfer reactions. The resulting peak in N$_2$H$^+$ abundance
outside of the CO snow line\footnote{As noted above, in order to reduce the number of
  free parameters, we put all CO in CO ices initially. This leads
  to an overestimation of the CO depletion in the less dense outer
  regions of the cloud. In order to check the influence of our initial
  conditions, we also calculated the model with one-quarter of
  the CO in the gas phase initially. The computation confirmed that
  while the CO depletion in the outer regions of the cloud is reduced
  by some orders of magnitude compared to the model presented here,
  the qualitative behavior of the chemical profiles does not change.}
at $\sim$70 au occurs at a radius of about
100 au, corresponding to 0.7$''$ at a distance of 140
pc. Accordingly, and shown by the synthetic observations created by the
\texttt{RATRAN} code and the \texttt{MAPPING} software, this ring would not be
resolved by our observations at 3 mm and would merely appear as
centrally peaked towards the source as shown in Fig. \ref{model1}.

The plateau abundance of C$^{18}$O is reached within the innermost 70
au. The resulting synthetic observations have a FWHM of 0.5$''$ in the
uv-plane as measured by Gaussian fits, which is about one-third the
size we find in our uv observations ($\sim$1.3$''$). Figure
\ref{model1} shows a comparison of the model and our observations both
as maps and as intensity cuts in the image plane. The apparent satisfactory
agreement in size of the modeled and the observed C$^{18}$O intensity
peaks is due to the fact that the modeled emission region is not
spatially resolved in the image plane and thus determined by the
synthetic beam size.

However this model, which is based on the temperature and
density profiles as derived from the present internal luminosity, is
not able to reproduce the location of the N$_2$H$^+$ emission peaks and
also deviates from the observed emission size of C$^{18}$O. The reason
for the first shortcoming is that the temperature at the radius of
$\sim$1300 au corresponding to the location of the observed N$_2$H$^+$
peaks is too low (only about 9.5 K, while $T_{\rm sub}$=19 K as found
in \citealt{Anderl:2016}) for the N$_2$ ices to be fully desorbed from the
grains. Furthermore, we do not observe centrally peaked N$_2$H$^+$
emission towards the source as predicted by the model.

\subsection{Modeling the N$_2$H$^+$ Peaks}
\label{burst}

\begin{figure*}
  \centering
  \includegraphics[width=17cm]{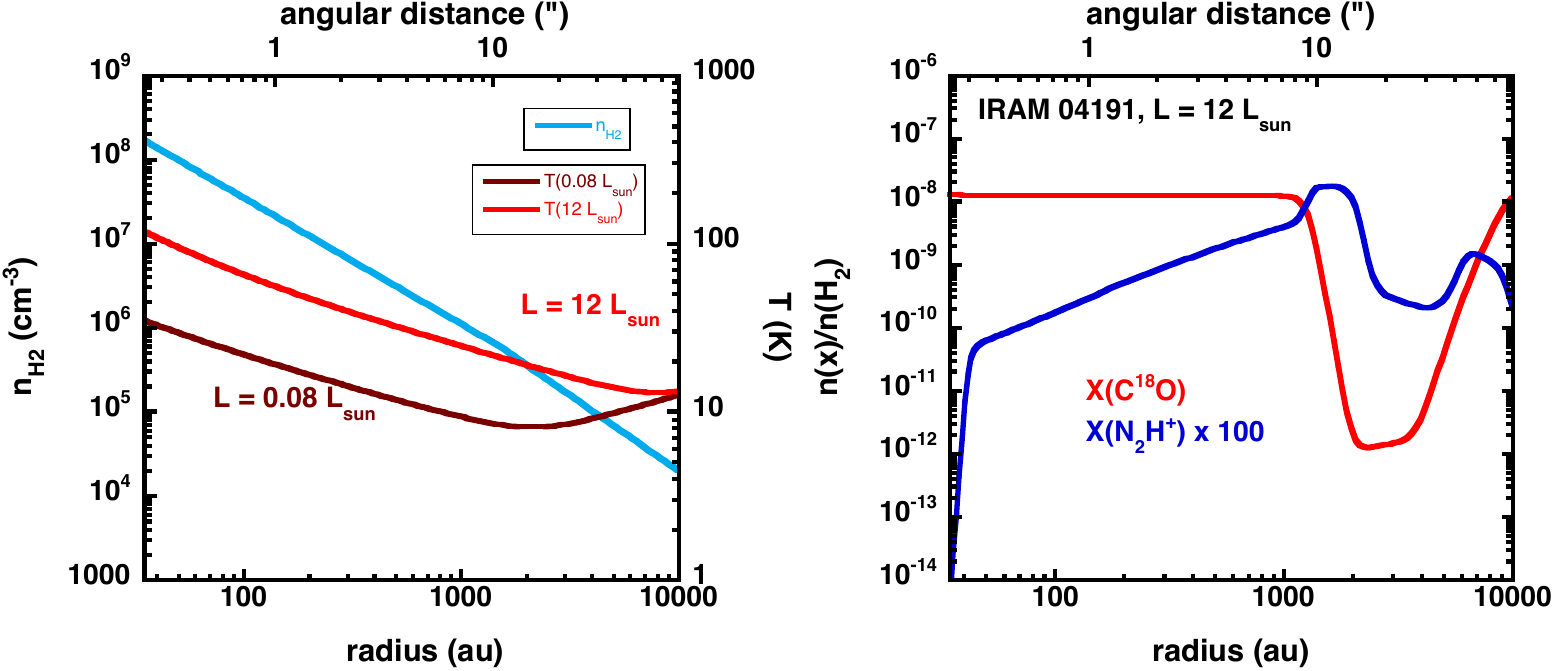}
  \caption{ Model with an increased source luminosity of 12
    L$_{\sun}$. {\it Left:} Density profile (light blue) and
    temperature profiles of IRAM~04191 for a source luminosity of 0.08
    L$_{\sun}$ (brown) and 12 L$_{\sun}$ (red). Both temperature
    profiles are self-consistently computed with the transphere
    code. {\it Right:} Chemical abundances of C$^{18}$O (red) and
    N$_2$H$^+$ (blue) calculated for a source luminosity of 12
    L$_{\sun}$, again after a chemical age of $\sim$10$^4$
    yr. \label{model2_prof_chem}}
\end{figure*}

The model with an internal luminosity of $L_{\rm int}$ $=$ 0.08~L$_{\sun}$
predicts a much smaller ring radius than the location of the observed N$_2$H$^+$ peaks. One possible conclusion could be to consider changing the N$_2$ binding energy used in the model. As already noted, we refrain from this possibility as the value used was found appropriate in all four sources analyzed in \citet{Anderl:2016} and is tightly constrained by laboratory studies (see the discussion in Section 5.4 therein).

Instead, we examine the possibility that the luminosity of the protostar may have been larger in the
past. To test this scenario, we attempt to constrain the luminosity
that would be required to shift the radius of the N$_2$H$^+$ peaks to
the observed distance. Using the \texttt{transphere} code to calculate
the temperature profile for the same density profile but an increased
source luminosity, we find that the radius of the central hole can be qualitatively reproduced if
the luminosity is increased by a factor 150, from 0.08 L$_{\sun}$ to 12
L$_{\sun}$ (we refer to this model as
model2). Figure~\ref{model2_prof_chem} shows the increased temperature
profile compared to the present day temperature profile together with
the resulting abundance profiles. The increased luminosity has now
shifted the N$_2$H$^+$ abundance peak out to radii of about 1500 au. This chemical abundance profile is already reached after 100 yr.

\begin{figure*}
  \sidecaption
  \includegraphics[width=12cm]{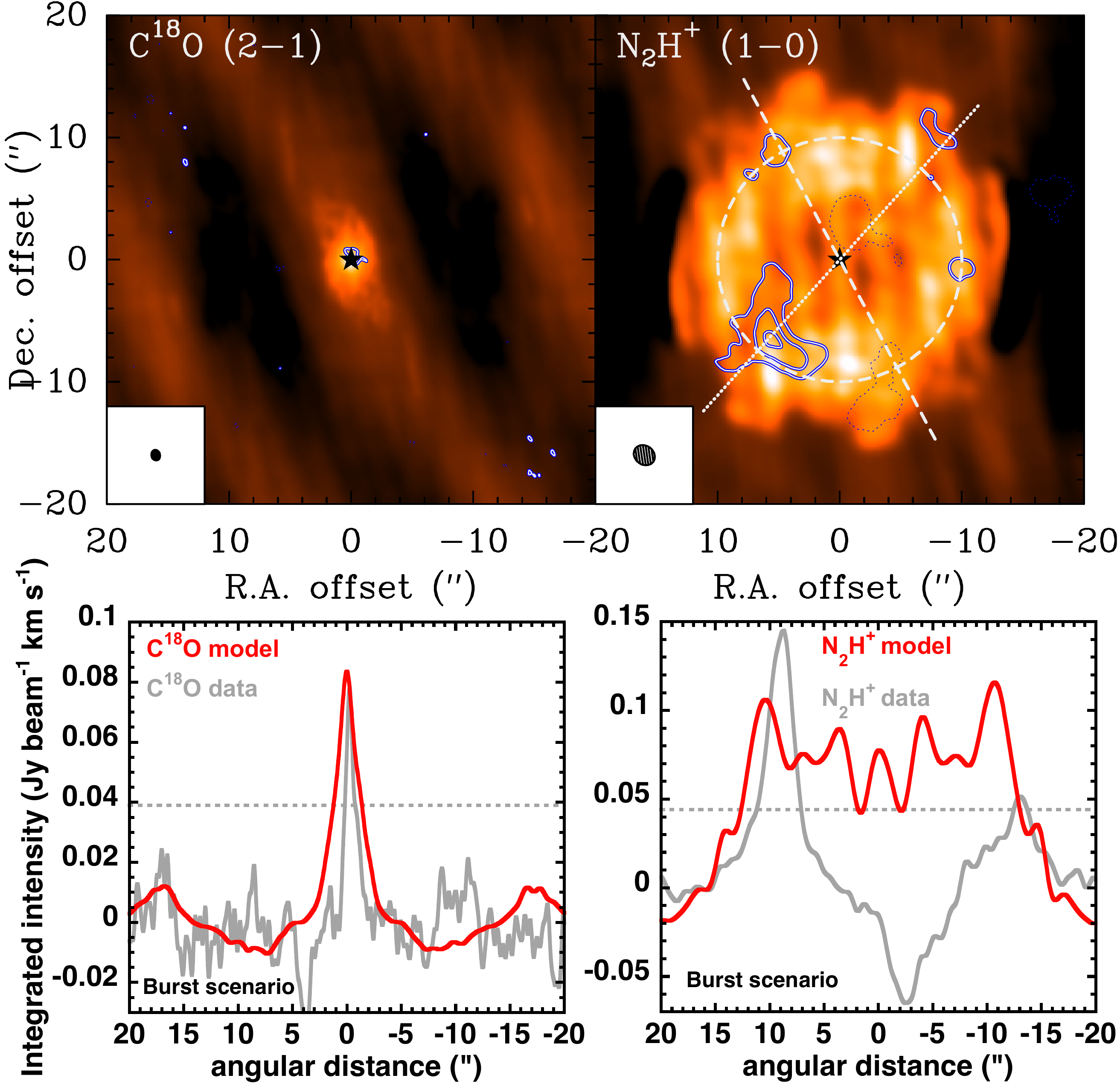}
  \caption{Same as Fig \ref{model1}, but for the model with increased
    source luminosity of 12 L$_{\sun}$. For the deconvolution of the
    maps, a mask enclosing the modeled emission region was used for
    the cleaning. \label{model2}}
\end{figure*}

Figure \ref{model2} shows the modelled data and the observed data in
the image domain for comparison. Despite differences in the detailed emission morphology, the location of the modeled
N$_2$H$^+$ peaks satisfactorily matches the observations. The modeled
C$^{18}$O central emission peak is slightly more extended than what we
observe (with an FWHM of $\sim$2.9$''$ it is about twice the size of
our observations). However, interestingly this emission region is not as large as the region inside of the N$_2$H$^+$ peaks in the image domain, even
though the plateau of high C$^{18}$O abundance reaches out to a radius
of $\sim$1100 au (Fig.~\ref{model2_prof_chem}). 

The cut of C$^{18}$O emission shown in Figure \ref{model2} is affected by
negative emission at radii further than 5$''$ from the source, similar
to what we see in our observations. Nevertheless, we note that also for a
cut in the vertical direction, which does not show negative emission
at radii inside the N$_2$H$^+$ hole, the C$^{18}$O emission between
angular radii of 5$''$ and 10$''$ is only weak, albeit positive (see
Figure \ref{time_all} for cuts in that direction). If this morphology
in the image plane is indeed reliable, it demonstrates that the
observed C$^{18}$O emission in this source does not directly trace the
full extent of the region where C$^{18}$O is desorbed from the grains.
The effect that the observed emission region can be substantially smaller than the true CO snow line radius 
was already found and discussed as one important result of \citet{Anderl:2016}.

\subsection{Time dependent modeling}
\label{timedep}

So far we have seen that (a) a constant low-luminosity model cannot
reproduce the observed N$_2$H$^+$ peaks and (b) a luminosity burst that
increases the present source luminosity by a factor of 150
would be able to release N$_2$ from the grains at the observed
radius. As a final step, we want to investigate 
whether or not the observed morphology in C$^{18}$O and N$_2$H$^+$ could be
the aftermath of a past luminosity burst. In order to study this, we
now use \texttt{astrochem} with a time-dependent
temperature profile. We start with the profiles from model1 and let it
evolve for 10$^4$ yr (we note that  this duration is not a crucial parameter); we then switch to the profiles from model2
(high-luminosity model) for 100 yr \citep[which is the typical burst
timescale; see e.g., ][]{Vorobyov:2013,Kuffmeier:2018}. We then
change the temperature profile to that of model1 again (low-luminosity
model), and let it evolve for some time, which is left as a free
parameter of the study. Our input parameters are listed in Table~\ref{modelparam}.

\begin{figure*}
  \centering
  \includegraphics[width=17cm]{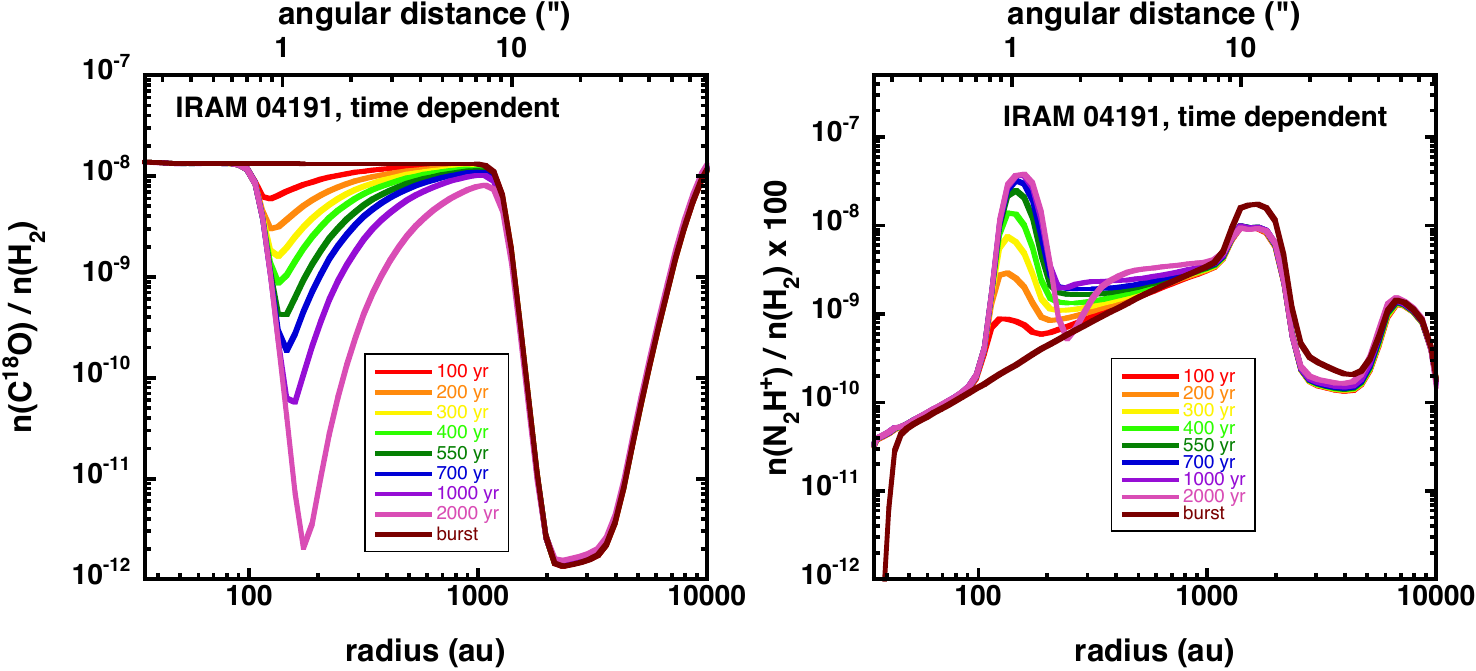}
  \caption{Chemical abundance profiles for different times after the
    end of the burst. {\it Left:} Chemical abundances of C$^{18}$O for
    nine different times up to 2000 yr after the burst. {\it
      Right:} Same for N$_2$H$^+$. \label{chemabun_time}}
\end{figure*}

Figure \ref{chemabun_time} shows the chemical evolution of the
C$^{18}$O and N$_2$H$^+$ abundance profiles in the envelope at the end
of the burst and 100, 200, 300, 400, 550, 700, 1000, and 2000 yr
after the luminosity has decreased back to its present low value. The
evolution starts with the abundance profiles created by the burst, as
discussed in Section \ref{burst}, where C$^{18}$O is found in the gas
phase inside a region with a radius of $\sim$ 1000 au and where
N$_2$H$^+$ peaks at a radius of $\sim$1500 au. Already after 100
yr, CO starts getting depleted just outside the new CO snow line
radius of $\sim$100 au, where the density is high ($\sim$7$\times$10$^7$~cm$^{-3}$) and where the
temperature is now low enough for CO to freeze out on the
grains (in our previous study we found the CO freeze-temperature to be $\sim$24 K; the CO freeze-out timescale for a density $\sim$7$\times$10$^7$~cm$^{-3}$ is $\sim$70~yr;  \citealt{Bergin:2007}). Accordingly, the destruction of N$_2$H$^+$ is halted at that
radius and a second N$_2$H$^+$ ring just outside the new CO snow line
emerges, while the peaks outside the old CO snow line are still visible
as a second local maximum. Within the next hundred years, the
depletion of CO proceeds further, moving outwards into regions of
lower density. The same is true for N$_2$H$^+$ in region of
temperatures lower than $\sim$19 K, decreasing the relative strength
of the outer N$_2$H$^+$ emission peaks with respect to the small, inner
ring. After 400 yr, the N$_2$H$^+$ abundance just outside the new
CO snow line at $\sim$140 au has become higher than the abundance at
the burst location of the snow line. For even later times, N$_2$H$^+$  is reduced even further at the location of the burst CO snow 
line.

\begin{figure*}
  \sidecaption
  \includegraphics[width=12cm]{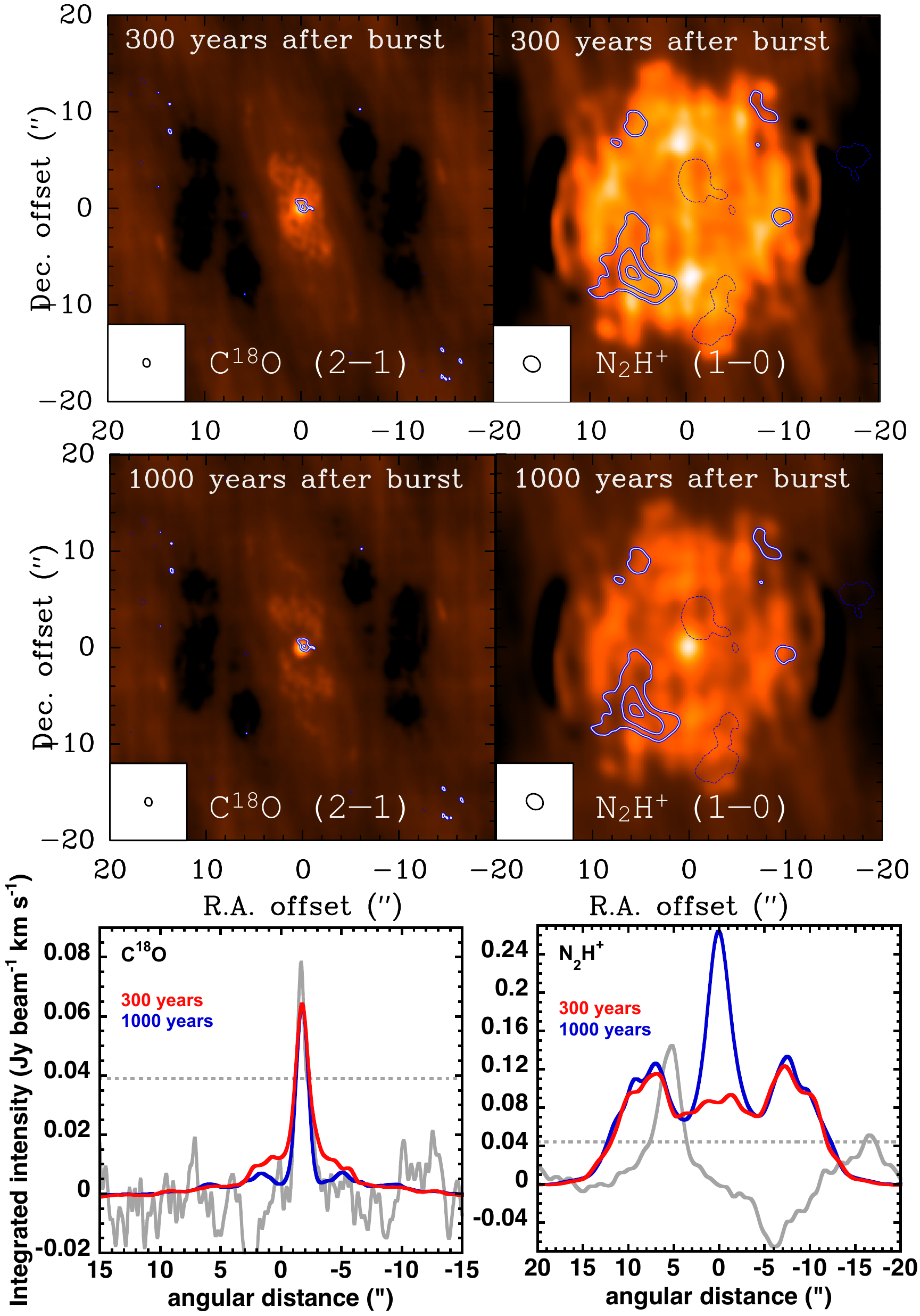}
  \caption{Top and middle panels: C$^{18}$O (2--1) (left) and N$_2$H$^+$ (1--0) emission (right) according to the time dependent
    model, 300 yr (top panel) and 1000 yr (middle panel) after the
    burst. Both panels show comparisons of the observations (contours) with the
    synthetic maps based on the respective models (color background).
    Integration intervals and the
    contour spacing are the same as in Fig. 1. The scaling of the maps
    can be understood in comparison with the cuts in the bottom panel. Bottom panel: Vertical cuts for C$^{18}$O (left) and N$_2$H$^+$ (right)  300 (red lines) and 1000 yr (blue lines) after the burst. Gray lines show the data. \label{time_all}}
\end{figure*}

Already from these N$_2$H$^+$ abundance profiles we may conclude that
if the burst scenario is correct, the observations hint at a past
burst that is not more than a few hundred years old, otherwise the
observed emission would likely be dominated by the new innermost N$_2$H$^+$
abundance peak. This conclusion is illustrated in Figure
\ref{time_all} (right panels), which show the
simulated observations in N$_2$H$^+$ 300 and 1000 yr after the
burst. While 300 yr after the burst, the emission still peaks at
the radius of the old snow line at $\sim$10$''$, 1000 yr after the
burst the central emission towards the source is more than twice as
intense as the emission at the burst snow line. However, a precise time estimate depends on the correct modeling of the N$_2$H$^+$ destruction process and is therefore prone to some uncertainty. If there is some additional destruction of N$_2$H$^+$ in the central part of the envelope we are not accounting for (such as destruction  by  CO$_2$ via  proton  transfer or by free electrons created by UV radiation or X-rays in the innermost envelope), as we hypothesized in \citet{Anderl:2016}, the time elapsed since the outburst could also be longer.

The comparison of our observations with the abundance profiles of
C$^{18}$O yields an opposite trend in principle (see Figure
\ref{time_all}, left panel): Right after the
burst, there is still some extended pedestal that illustrates the
presence of C$^{18}$O in the gas phase outside of the present snow
line. This pedestal has largely vanished after 1000 yr. Given the
low signal-to-noise ratio of our C$^{18}$O observations. However, we
consider both models to be consistent with our observations in
C$^{18}$O and hence N$_2$H$^+$ being the better tracer of the time
passed after the burst. That being said, from our time-dependent
modeling it appears that if a burst is indeed responsible for the
ring of N$_2$H$^+$ emission at radii of $\sim$10$''$, the burst
should not have occurred much more than a few hundred years ago.

In summary, reproducing our observations seems to require a trade-off
between two counteracting processes: On the one hand, enough time
should have passed after the burst for the C$^{18}$O abundance outside
the present CO snow line to have been sufficiently decreased by CO
freeze-out in order to explain the observed compact C$^{18}$O peak. On
the other hand, the burst should have been close enough in time for
the outer N$_2$H$^+$ emission peaks to still dominate the emission. This second
requirement might be slightly relaxed if there is a process in the
innermost envelope that destroys N$_2$H$^+$, which is not included in
our model. The existence of such a mechanism was already hypothesized
by \citet{Anderl:2016}, who suggested
either destruction by CO$_2$ or by free
electrons. Such a mechanism could be the cause of a lack of a central maximum in N$_2$H$^+$ emission. However,
the requirement to see clear emission peaks in N$_2$H$^+$ without
any central emission still seems to make after-burst ages of much more
than 1000 yr unlikely. We conclude that the time-dependent models
at ages of not more than 1000 yr appear to be qualitatively consistent with
our observations.

\section{Discussion}
\label{Discussion}
\subsection{Evidence for a past accretion burst}
\label{Discussion_1}

As we see in our analysis, our observations yield strong
evidence for a past accretion burst in IRAM 04191: The
central hole of N$_2$H$^+$ emission (as already observed
by \citealt{Belloche:2004} and \citealt{Lee:2005}) cannot be reproduced with a chemical
model based on the present source luminosity and the resulting
temperature profile. However, the observed emission peaks 
satisfactorily match our model, which assumes a past episode
of increased luminosity. This model also accounts for the compact
emission component we see in C$^{18}$O. The first result in particular seems very solid: At the radius where
N$_2$H$^+$ is observed, the temperature profile corresponding to the
present protostellar luminosity is not high enough to allow for
N$_2$H$^+$ to be present in the gas phase. 

In addition, there might be another direct way to probe our
hypothesis of a past accretion burst. \citet{Vorobyov:2018} point out that protostellar jets,
in addition to chemical signatures of past accretion burst, can provide
important evidence for the history of protostellar accretion \citep[see also][]{Froebrich:2016}. Building
on the study by \citet{Arce:2007}, \citet{Vorobyov:2018} use hydrodynamical simulations
to show how the time spacings between luminosity bursts and the
knotty protostellar jets in CARMA 7 relate to each other. From their
general agreement between model and observations, they infer a causal
link between episodic mass accretion and the knotty jet
structure. This latter
study seems to suggest that chemical analyses of
accretion burst can and probably should use information from the
observed jet structure as a complement. This begs the question of whether or not there any hints for our
hypothesis of a past accretion burst in the outflow of IRAM 04191.

There is some additional evidence to be considered. \citet{Lee:2005}  mapped the emission around IRAM 04191 in
several molecular lines with the BIMA array and traced a bipolar
outflow in CO, HCO$^+$, and CS. These latter authors detected two internal structures:
an internal bow shock visible in HCO$^+$ and CO $\sim$20$\arcsec$
(2800 au) northeast of the source, and a linear HCO$^+$ structure of
$\sim$50$\arcsec$ length southwest of the source. The former feature
is attributed to a presumptive temporal variation in the jet
velocity. Assuming an inclination angle of the outflow axis to the
line of sight of $\sim$50$^\circ$ \citep{Belloche:2002} together
with an assumed jet velocity\footnote{\citet{Lee:2005} use a jet-driven bow-shock model to interpret their observations. This model suggests a bow shock advance speed of $\sim$25 km s$^{-1}$. However, typical jet velocities as prompted by MHD simulations are higher: up to 100 km s$^{-1}$.} of 25-100 km s$^{-1}$, the age of that
structure would amount to about 200-825 yr. This value matches the
result of our chemical analysis very well, and suggests that the burst should have occurred not more than 1000 yr ago. This  nicely demonstrates the potential synergy of both methods ---the outflow analysis and the modeling of snow line chemistry--- in the analysis of protostellar accretion history.

\subsection{Does infall have an impact?}
\label{Discussion_2}
Our model
does not account for the infall of envelope
material, and so we need to consider whether or not this shortcoming indeed has an impact on our analysis. Some simple estimates can shed  light on this question: \citet{Belloche:2002} determined the infall velocity at a
radius of 1150 au as being smaller than 0.2 km
s$^{-1}$. \citet{Lee:2005} measured the infall velocity at the peak
radius as vanishing. If we take the maximum
value suggested by Belloche et al., the
material at the peak location would move by  approximately 40 au in 1000 yr. As the physical
conditions do not significantly change at such small distances, we
believe that our static stationary model does not jeopardize our
conclusion. The same holds for our model with time-dependent physical conditions, as the
considered timescales are very short. In summary, modeling the infall motion of the gas does not seem to be significant for the case studied.

\subsection{Fate of IRAM 04191}

\citet{Hsieh:2018} performed a similar study to ours with a sample
of seven VeLLOs observed in CO isotopologs and N$_2$H$^+$ with
ALMA. Applying a very similar modeling procedure to the one
introduced in this paper (although without not producing synthetic observations), they evaluated whether the position of N$_2$H$^+$ peak emission is consistent with the current source luminosities, and found that five of their sources showed
chemical abundance profiles that pointed at past accretion bursts: their peaks were significantly further apart from the central source than explained by the assumed current luminosity.

Our study adds one more VeLLO source to this list of five possible post-burst sources. Furthermore, the burst properties that we find here very closely match those of \citet{Hsieh:2018}: Our burst luminosity of 12 L$_{\sun}$ lies within the range of values between 1.0 and 30 L$_{\sun}$ found by these latter authors, and also our burst CO snow line radius of 1100 au matches their results with values between 600 and 2500 au. To further compare our findings with theirs, we can also compute the mass accretion rate during the burst based on the assumption that the burst luminosity is equal to the accretion luminosity $L_{\rm acc}$=$L_{\rm burst}$. For this, \citet{Hsieh:2018} use the following expression: $L_{\rm acc} = G \, M_{\rm star} \, \dot{M_{\rm acc}} / R$ where G is the gravitational constant, R is the protostellar radius (which they assume to be 3 R$_{\sun}$), and $M_{\rm star}$ is the mass of the central source (which they assume to be 40 M$_{\rm Jupiter}$). Using this equation together with the values for $R$ and $M_{\rm star}$ that they assume, we derive a burst mass accretion rate for IRAM 04191 of 3$\times$10$^{-5}$ M$_{\sun}$~yr$^{-1}$. This, again lies well within their range of values, namely between 6$\times$10$^{-6}$ and 4$\times$10$^{-5}$ M$_{\sun}$~yr$^{-1}$.

Now we can use this value to derive an estimate for the mass IRAM 04191 will be able to reach by the end of the Class 0 phase in this scenario of intermittent bursts. \citet{Hsieh:2018} derive a timescale of intervals between bursts in VeLLOS between 12,000 and 14,000~yr. If the burst itself lasts for $\sim$100~yr \citep{Vorobyov:2013}, the VeLLO will spend about 1\%\ of its lifetime ---which we can assume to be (0.4 - 0.5)~$\times$10$^6$~yr \citep{Evans:2009}--- in a burst state. The mass-accretion rate for IRAM 04191 in its quiescent phase, characterized by an internal luminosity of $L_{\rm int}$=0.08 L$_{\sun}$, is 2$\times$10$^{-7}$ M$_{\sun}$ yr$^{-1}$. Taking all this together, the final mass of IRAM 04191 will be 0.2 - 0.25 M$_{\sun}$, which is larger than the stellar/substellar limit. Indeed, if our scenario of a past accretion burst together with the assumed timescales is correct, the VeLLO IRAM 04191 should accrete enough mass to become a low-mass star rather than a brown dwarf.

 \section{Conclusions}
\label{Conclusions}

As part of the CALYPSO IRAM Large Program, we observed C$^{18}$O
and N$_2$H$^+$ towards the Class 0 protostar IRAM 04191 with the IRAM
Plateau de Bure interferometer at sub-arcsecond resolution. Our
conclusions can be summarized as follows:

\begin{itemize}

\item Our CALYPSO PdBI data confirm previous observations of a central hole in the
  emission of N$_2$H$^+$, with an approximate radius of
  $\sim$10 $''$ around the continuum source. The C$^{18}$O emission is
  centrally peaked with an FWHM perpendicular to the outflow axis of
  1.3$''$. This very compact peak is still visible if the
  interferometer data are combined with the emission on larger scales,
  as observed with the IRAM 30m telescope. This is in contrast with
  other Class 0 protostars, in which the C$^{18}$O emission typically
  fills an N$_2$H$^+$ emission ring \citep{Anderl:2016}.

\item With a model based on the present source internal luminosity of 0.08
  L$_{\sun}$ we can only reproduce the observed centrally peaked
  emission in C$^{18}$O. We are unable to reproduce the observed
  emission in N$_2$H$^+$. The observed emission peaks in N$_2$H$^+$ are
  nevertheless consistent with a model based on a luminosity of about 12
  L$_{\sun}$.

\item Using a time-dependent chemical model that implements a past
  burst episode, we can model the observed emission morphology in
  N$_2$H$^+$ and also in C$^{18}$O  to a satisfactory degree. Based on
  this model, we can constrain the time of the burst to no longer
  than 1000 yr ago.

 \item Comparing our results with those of \citet{Hsieh:2018}, our burst luminosity, radius of the burst snow line, and burst mass accretion rate are similar to the values they found for five post-burst candidates in a set of seven sources. Using their estimate for the timescale of intervals between the bursts, we derive an expected final mass of IRAM 04191 of 0.2 - 0.25 M$_{\sun}$: If the burst scenario is correct, IRAM 04191 will reach a mass  higher than the typical hydrogen-burning limit mass and thus become a low-mass star at the end of the Class 0 phase (while it will continue accreting mass during the subsequent Class I phase).
 
\end{itemize}
 
Our analysis shows that the combination of N$_2$H$^+$ and C$^{18}$O, rather than the use of one of these two molecules alone, can be used as a powerful tracer of potential past luminosity outbursts. As our modeling demonstrates, the respective chemical evolution after a burst yields tight constraints for the time that has passed since the last burst. In particular, for IRAM 04191, comparatively little time must have past because the peaks of N$_2$H$^+$ emission are still very clearly visible as compared to the central compact emission, which stems from the location outside the post-burst CO snow line.

 \begin{acknowledgements}

   We would like to thank an anonymous referee for constructive and valuable comments that helped to strengthen the paper. This work is based on observations carried out under project number
   U052 with the IRAM NOEMA Interferometer. IRAM is supported by
   INSU/CNRS (France), MPG (Germany) and IGN (Spain). This work has
   benefited from the support of the European Research Council under
   the European Union?s Seventh Framework Programme (Advanced Grant
   ORISTARS with grant agreement No. 291294 and Starting Grant
   MagneticYSOs with grant agreement No. 679937), and from the French
   Agence Nationale de la Recherche (ANR), under reference
   ANR-12-JS05-0005. 

 \end{acknowledgements} 
 
   \bibliographystyle{aa}
\bibliography{biblio}


\begin{appendix}

\section{Shortspacing correction}
\label{shortspacing}

\begin{figure}
  \includegraphics[width=\columnwidth]{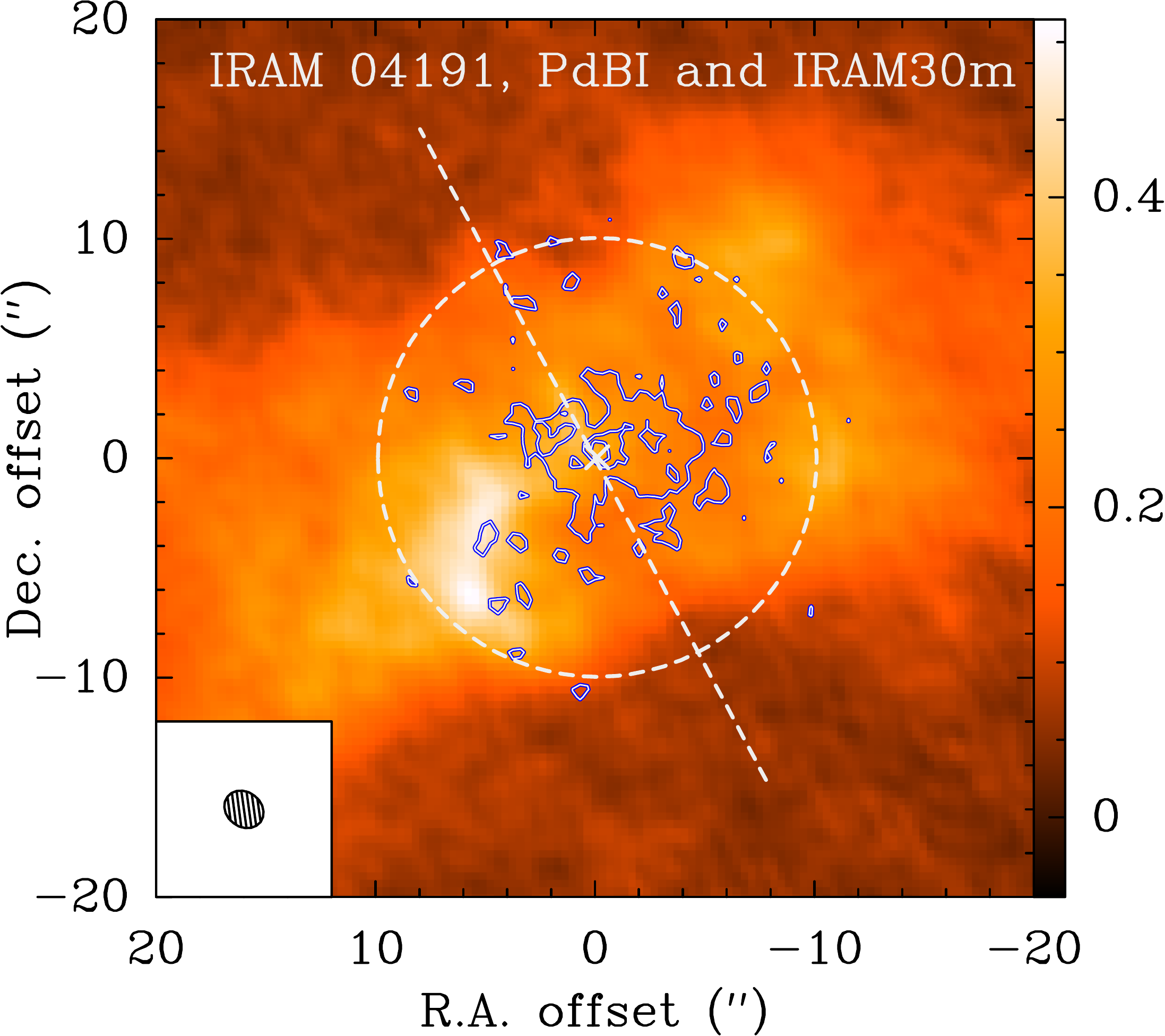}
  \caption{Combined PdBI and 30m
    data: N$_2$H$^+$ (1--0) emission integrated over all seven hyperfine components is shown as colored background. The integrated N$_2$H$^+$ map has an RMS noise value of
    $\sigma$ = 0.028 Jy beam$^{-1}$ km s$^{-1}$. Contours show integrated C$^{18}$O (2--1) emission, starting from
    6$\sigma$ in steps of 3$\sigma$ with $\sigma$ = 0.021 Jy
    beam$^{-1}$ km s$^{-1}$. The merging was done with the relative
    weighting automatically suggested by the GILDAS
    software. \label{map_shortspacing}}
\end{figure}

\begin{figure}
  \includegraphics[width=\columnwidth]{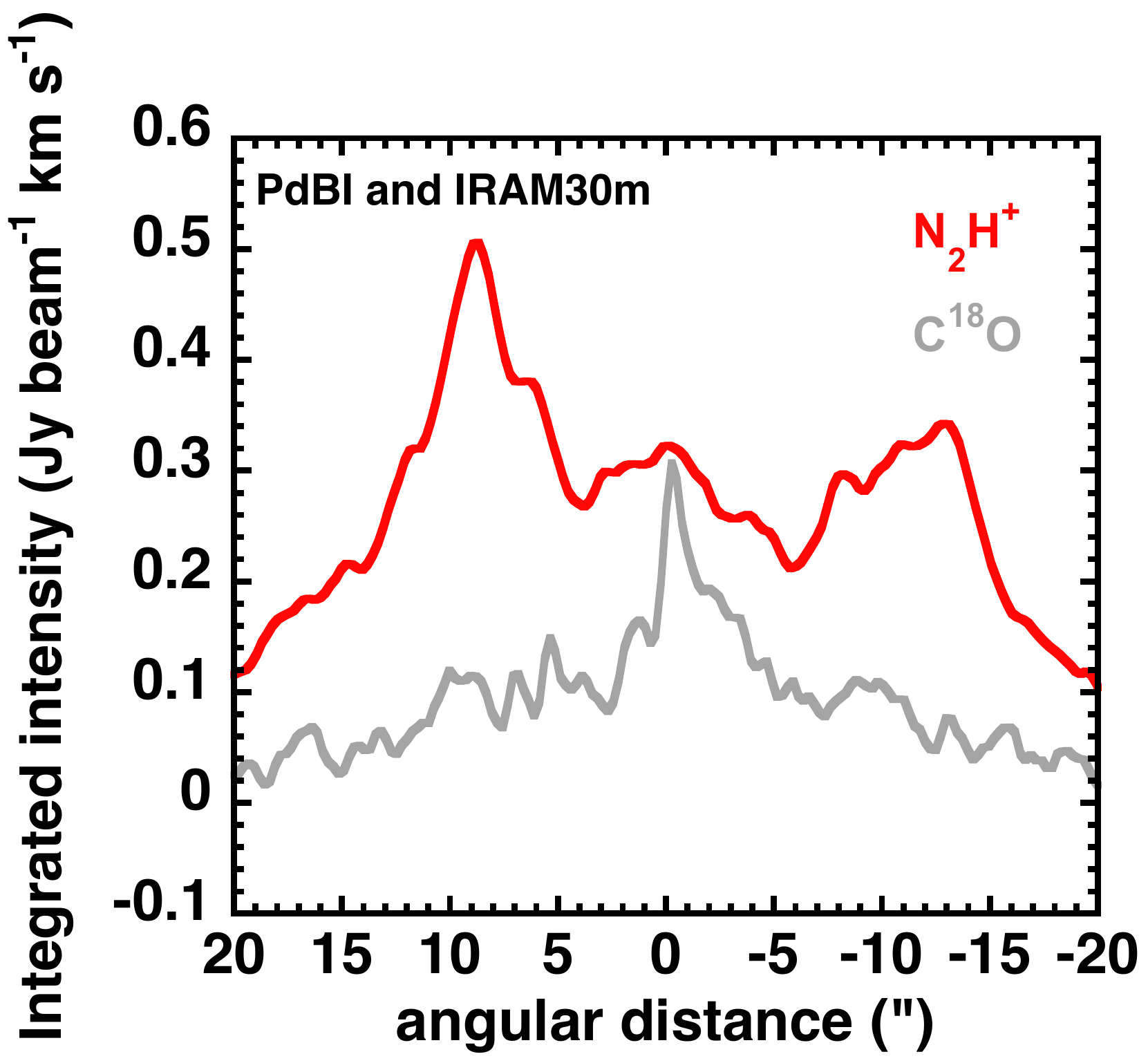}
  \caption{Same as Figure 2, but for the combined PdBI and 30m
    data. \label{cuts_shortspacing}}
\end{figure}

The approximate radius of the observed N$_2$H$^+$ rim in the PdBI
data is 10$''$. This is at the edge of the PdBI interferometer's 
primary beam at the frequency of the C$^{18}$O
(2--1) emission. 
We therefore have to check the possibility that more extended
emission in C$^{18}$O than what we see in the PdBI data is simply filtered
out. Figure \ref{map_shortspacing}. shows the combined PdBI and 30m
data. The details of how the short-spacing correction was performed can be found in \citet{Gaudel:2020}.

The morphology of the N$_2$H$^+$ rim as seen in the PdBI data
alone is still clearly visible on top of the flattened envelope that
was already observed by \citet{Belloche:2002} and
\citet{Lee:2005}. Indeed, there is a more extended, broad component of
emission in C$^{18}$O, which does not seem anticorrelated with the
emission in N$_2$H$^+$. The intensity cuts shown in Figure
\ref{cuts_shortspacing} show that the narrow peak of
C$^{18}$O emission that is seen in the PdBI data is still visible in
the combined data. We therefore assume that the extended component stems from the envelope while the central sublimation region is
traced by the compact emission peak. In summary, we conclude that the two key characteristics of the data that we base our analysis on, namely the location of the N$_2$H$^+$ emission peaks and the size of the region of enhanced C$^{18}$O emission, do not depend on the inclusion of the short-spacing data. In particular, the validity of our analysis will not depend on the specific N$_2$H$^+$ abundance inside the rim.

\section{N$_2$H$^+$ and C$^{18}$O spectra}
\label{sec:n_2h+-c18o-spectra}

\begin{figure}
  \includegraphics[width=\columnwidth]{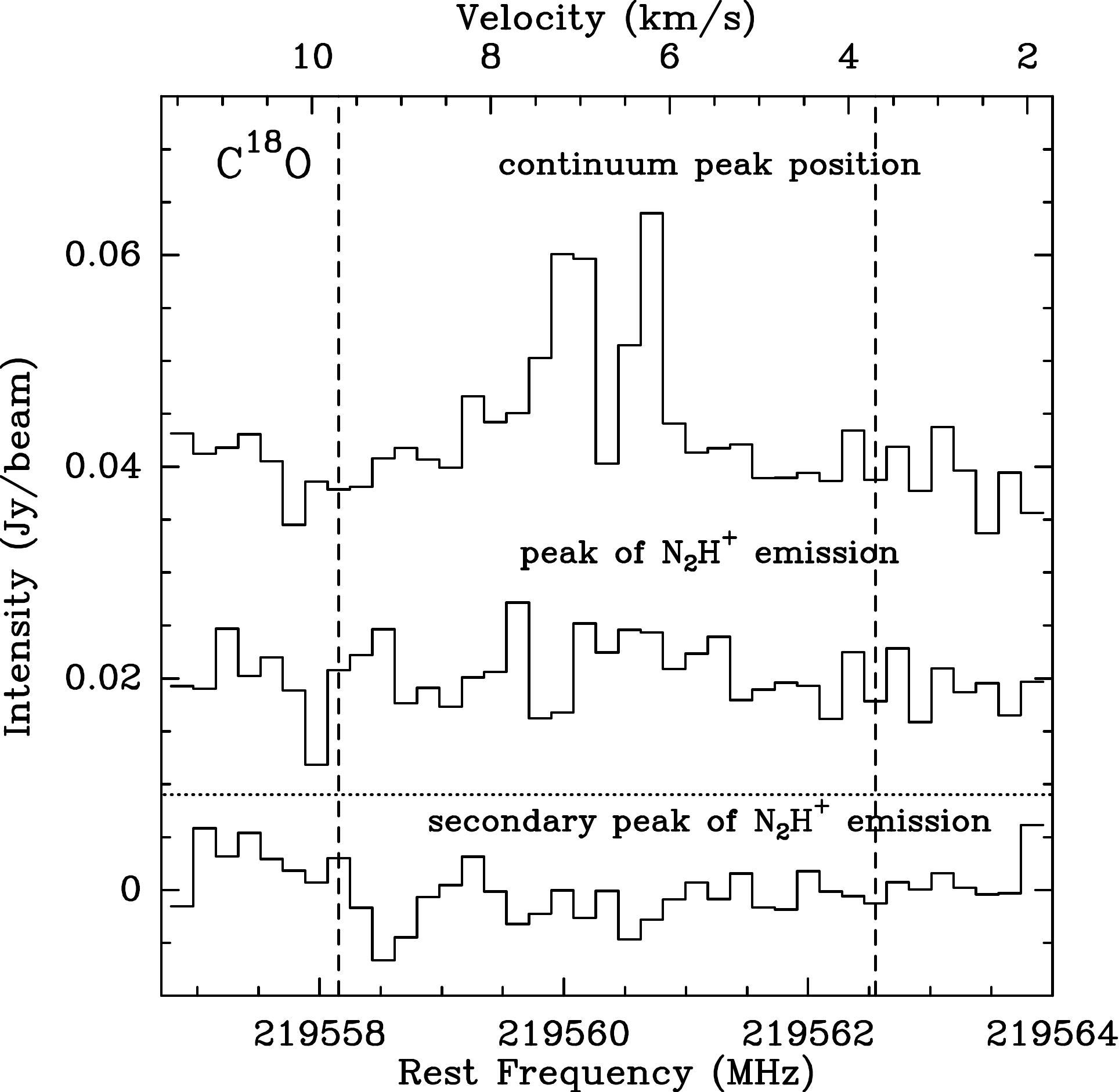}
  \caption{Spectra of C$^{18}$O (2--1) at the continuum peak position
    (top) and at the main (middle) and secondary peaks (bottom) of
    N$_2$H$^+$ emission. The top and middle spectra are
    shifted vertically for clarity. The dashed lines indicate the
    velocity interval used for computing the integrated intensity. The
    dotted line shows the $3\sigma$ level ($1 \sigma = 3.0$~mJy~beam$^{-1}$). 
    \label{c18o_spectra}}
\end{figure}

\begin{figure*}
  \sidecaption
  \includegraphics[width=12cm]{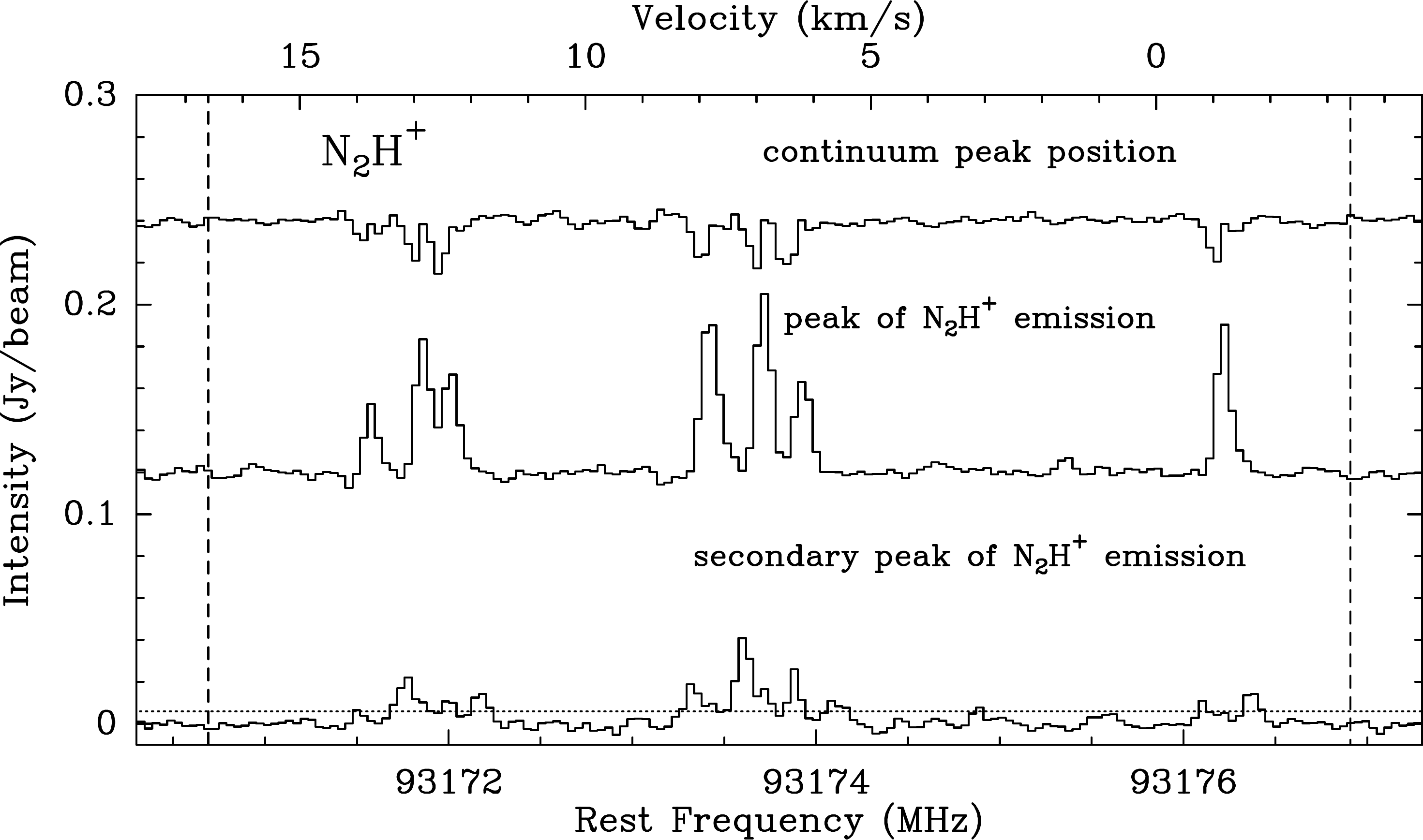}
  \caption{As Figure \ref{c18o_spectra} but for the seven hyperfine
    structure components of N$_2$H$^+$ (1--0). The dotted line
      shows the $3\sigma$ level ($1 \sigma = 2.0$~mJy~beam$^{-1}$).
    \label{n2h_allspectra}}
\end{figure*}

While for our analysis we only work with the integrated
  spectral line emission (C$^{18}$O 2--1 at 219.560\,354 GHz is
  integrated over $\pm$3 km s$^{-1}$ around the systemic velocity and
  the seven N$_2$H$^+$ hyperfine components of the 1--0 transition
  over a window of 20 km s$^{-1}$), it may still be revealing to see
  the shape of the spectra on different locations. Figure
  \ref{c18o_spectra} and \ref{n2h_allspectra} show the respective
  molecular emission in three locations: the continuum peak position
  and the main and the secondary peaks of the N$_2$H$^+$ emission.  By
  fitting the intensities of all hyperfine structure (HFS) components
  towards the N$_2$H$^+$ peak position with CLASS, we find that the
  opacity at the line center of the each HFS component is comprised
  between 0.3 and 2. Therefore, these components are optically thin to
  moderately optically thick.  We note that no lines  other than the
  N$_2$H$^+$ HFS components are detected in the 20 km s$^{-1}$
  window. We also checked in the CDMS database \citep{Muller2005} that
  no other bright lines are expected in this frequency window. It is
  therefore unlikely that the integrated emission is contaminated by
  other lines.

\end{appendix}

\end{document}